# Texture transitions in the liquid crystalline alkyloxybenzoic acid 6OBAC

A. Sparavigna[1], A. Mello[1], and B. Montrucchio[2]
[1] Dipartimento di Fisica, Politecnico di Torino
[2] Dipartimento di Automatica ed Informatica, Politecnico di Torino
C.so Duca degli Abruzzi 24, Torino, Italy

The 4,n-alkyloxybenzoic acid 6OBAC has a very rich variety of crystalline structures and two nematic sub-phases, characterised by different textures. It is a material belonging to a family of liquid crystals formed by hydrogen bonded molecules, the 4,n-alkyloxybenzoic acids  $n$  indicates the homologue number). The homologues with $7 \leq n \leq 13$ display both smectic C and N phases. In spite of the absence of a smectic phase, 6OBAC exhibits two sub-phases with different textures, as it happens in other materials of the homologue series which possess the smectic phase. This is the first material that exhibits a texture transition in a nematic phase directly originated from a crystal phase. Here we present the results of an image processing assisted optical investigation to characterise the textures and the transitions between textures. This processing is necessary to discriminate between crystal modifications and nematic sub-phases.

**Introduction**
The mesogenic features of the alkyloxybenzoic acids are due to the presence of hydrogen-bonded dimers, where the strength and the orientation of the hydrogen bonds provide the conditions for the presence of smectic and nematic phases. Peculiar behaviours, like textures transitions in the nematic phase and the appearance of domains of spontaneous twist and dendrite structures [1-5], makes liquid crystals with hydrogen bonded dimers, such as 4,n-alkyloxybenzoic acids, very attracting materials for the research of chiral behaviour in non-chiral compounds. In the nematic phase, the material is a mixture of monomers and dimers (closed and open, Fig. 1), with concentrations depending on the temperature. Near the nematic-isotropic phase transition, as reported in [4], a process of oligomerization, with the formation of trimers, quadromers and so on, is possible: these oligomers produce a chiral nematic domain in a material with achiral molecules.

In the 4,n-alkyloxybenzoic acids (nOBAC), the monomers are composed of two sterically distinct molecular parts, the oxybenzoic acid residue and the aliphatic chain. The number $n$ of carbon atoms in the aliphatic tail gives origin to the homologue  nOBAC series. The molecule unit determining the behaviour of the member in the series is the acid residue when the chain is short: no mesophases are present in the first and in the second member of the series 1OBAC and 2OBAC, where the melting point is above the temperature range of mesophase stability. From 3 to 6 carbon atoms in the tail, nematic but no smectic phase is present. Acids from 7 to 18 carbon atoms in the alkyl tail are smectogenic. The homologues with $7 \leq n \leq 13$ display both smectic C and N phases.

Optical and dielectric investigations of some of them (7OBAC, 8OBAC, 9OBAC) show a nematic phase affected by two different ordering at low and high temperature. This was thought as due to the presence of cybotactic clusters [6] having short-range smectic order in the nematic phase under a certain temperature. In the nematic melt, if the temperature is low enough, dimers aggregate in clusters with smectic C ordering [6-11].

The member with 6 carbon atoms in the tail, 6OBAC (monomer $C_6H_{13}OC_6H_{10}COOH$) is not able to display a smectic phase. Nevertheless it is a quite interesting material. We have found by means of the optical observations that, in spite of the absence of a smectic phase, the nematic phase has two sub-phases, like 7-, 8- and 9OBAC. Moreover, this material has a rich variety of crystals modifications. And in the nematic phase, its behaviour is resembling that of the hexylcyclohexane carboxylic C6 acid, material dimerizing via hydrogen bonds too [12,13]

**Image processing assisted optical investigations of nematic and crystal phases.**
The liquid crystal cell used for optical investigation was of a common sandwich type, with no spaces between glasses. The liquid crystals was heated in the isotropic phase and introduced by capillarity in the cell. The surface of the glasses was previously rubbed to favour the planar alignment. The cell, placed in a thermostat, was observed with a polarised light microscope connected with a digital image acquisition system. To identify the temperature transition between the nematic sub-phases, we used an image processing analysis previously applied to study 7-, 8OBAC and C6 [12,13]. In the appendix we give details of calculations. Four crystalline textures and two nematic sub-phases are observed with transition temperatures as in table I. Fig.2 shows the four crystal sub-phases. By a simple visual inspection, the four crystalline textures are difficult to distinguish, due to the slight changes from a crystal texture to the other: the image processing is then necessary to give a precise value of the transition temperature. The same is made for the study of the nematic phase. In the nematic range, two sub-phases (Fig.3) are displayed, similar to those observed in other members of the series (7-,8- and 9OBAC) and in the trans-4-hexylcyclohexane carboxylic acid (monomer $C_6H_{13}C_6H_{12}COOH$).

The solid state organisation of 4,n-alkyloxybenzoic acids was investigated by means of the X-ray equipment and the molecular reorganisation proposed by Bryan et al. [14-16]. The transition temperatures reported in Table I are in agreement with those in Ref.[16]: they observed in fact five crystal modifications but the last crystal-crystal transition is very close to the crystal-nematic transition and practically coalescent with it. This phase could be a smectic-like phase that is not able to develop itself, a sign of which we could find in the existence of the two nematic sub-phases.

Bryan et al. observed the absence of optical recording of crystal polymorphism in the previous reports on nOBAC series. In fact, to reveal this crystal polymorphism, the liquid crystal cell needs a long time to rest after preparation, to relax in the crystal form Cr-I. Immediately after the insertion of the isotropic liquid, the optical investigation of the cell reveals only one crystal texture below the nematic phase. Fig.2 shows the four crystal textures: the transition from Cr-I to Cr-II appears as a subtle wave enlarging from some domains in the cell frame and overcoming the entire frame. From Cr-II to Cr-III, the transition presents itself as white and black spots in the frame, but the largest apparent difference is gained in the transition from Cr-III to Cr-IV, where the sample assumes a final mosaic texture. We were not able to find the fifth crystal found in Ref.[16], although we lowered the temperature rate of the thermostat. According to [16], the presence of so many crystal phases is due to the difficulties of the molecule with an intermediate length of the alkyl tail to package themselves in a well defined crystal structure: changes in the temperature compel the molecules to try other ways in packaging.

The image processing previously used to distinguish smooth transitions is here applied to analyse the four crystalline textures. An image frame was recorded each time that the temperature increases of one degree. The colour image sequence was then analysed associating to each pixel at the point $(x,y)$ of the image frame, a colour tone (green, for instance), obtaining in this way a function representing the image colour intensity distribution $g(x,y)$ in the image frame. The mean value and the statistical moments of the colour distribution $g(x,y)$ are easy to evaluate (see Appendix for details). As it is possible to see in Fig.4, which proposes the mean value and variance of $g(x,y)$ as functions of the temperature. Mean value is not sensitive to transitions, but variance has discontinuities. What is much more sensitive to put in evidence the transition temperatures is the function obtained evaluating the difference between one image frame and the following in the scanning sequence: as the crystal texture changes, the difference has a peak. To check that it is not an artefact, we associate to each image frame a polar diagram giving the mean shape and size of domains in the image frame (see Appendix for details), with the same algorithm used in [12]. A sequence of polar diagrams (one representative for each of the four crystal textures is shown in Fig.5) confirms the texture variations at the transitions found in Fig. 4.

In the crystal texture Cr-IV, striped areas are visible (see fig.6). It is interesting to compare this texture with those observed in 8OBAC. X-ray investigations tell that this material must have two crystalline phases [16]. To observe these two crystalline modifications with microscope, the cell must rest for a long time after preparation, otherwise only one crystal phase is displayed by the material. The crystal texture preceding the smectic phase is the one which is not possible to observe immediately after preparing the cell. This phase is considered the precursor of the smectic phase. The precursor in 8OBAC has a texture with striped domains as Cr-IV in 6OBAC (see Fig.7). Cr-IV could be then considered a precursor, but not of a smectic phase, that is not observed. The alkyl tail is too short to have a smectic phase, nevertheless it is long enough to create cybotactic clusters that are responsible of the presence of two sub-phases in the nematic range.

The transition from the precursor Cr-IV into the nematic sub-phase N1 is shown in the Fig.8: it extends on a temperature interval of two degrees. From the images in Fig.8 it is possible to see traces of the crystalline texture in the nematic texture. To check the memory effect, we perform the characterisation of textures with a more detailed polar diagram, for shape and size evaluation of domains in the image frame. The shape of the domains, according to the procedure in Ref.[12], is the same for Cr-IV and nematic N1. The size is modified.

**Nematic phase.**

As already observed in 7-, 8- and 9OBAC, by means of dielectric investigations and optical inspections, the nematic phase exhibits a different order at low and high temperatures. The optical microscopy shows two textures then in the nematic range. In 6OBAC, as for 7OBAC, the texture transition does not correspond to any peak in differential scanning calorimeter

DSC measurements [11]. This is consistent with the hypothesis that it is the concentration of cybotactic clusters the origin of the transition. For 8OBAC and 9OBAC, it appears a small peak in the nematic phase in the calorimetric signal on heating and cooling the samples. To the authors' knowledge, 6OBAC is the first materials that shows the texture transition in a nematic phase coming from a crystal phase.

A model to explain the origin of the different textures in the nematic phase in proposed in the sequence in Fig.9. Let us suppose to start from a smectic C structure with schlieren defects, as it happens for instance in 7OBAC, inserted in a thin planar cell. The schlieren defect is formed by different orientations of the smectic planes with respect to the cell walls. After the transition in the nematic phase, smectic planes disappear but in some points they remain anchored at the walls and there, cybotactic clusters maintain the local smectic order. Defects on the walls help the persistence of these clusters. When the temperature further increases, the smectic order in cybotactic cluster is suppressed. The persistence of local smectic order at the walls explains the memory effect seen in Fig.8, if we suppose that the mechanism in the formation of the textures in 6OBAC is the same as in 7OBAC, even if the smectic phase does not grow.

Another interesting feature of 6OBAC is the following. In the sub-phase at high temperature N2, 15 degrees before the clearing point, it is observed an abrupt change in the colours of the domains in the nematic phase. In Fig.10, it is shown a domain of the nematic cell and how its colour passes from green to red and then to green again. On the right side of the figure, maps of the histograms for the three colours tones is proposed. This behaviour was previously observed in C6 [13], and it is not common in nematics. The reason of a change in colours of a cell viewed with polarised light is in the variation of ordinary and extraordinary indexes of the material. Let us remember that the material in the nematic phase is a mixture of closed, open dimers and monomers. Approaching the clearing point, the number of closed dimers decreases. One of the to hydrogen bonds forming the dimers is broken and the structure of the dimers is changed. Open dimers can connect with monomers to form trimers, or more complex oligomers, as in C6. The optical indexes are related to the nematic order which in turn is depending on the relative concentrations of dimers (closed and open), monomers and oligomers. The same behaviour of 6OBAC and C6 suggests that in 6OBAC too the formation of oligomers is possible and then of chiral domains, but we have not observed till now.

**Conclusions.**

The crystalline modifications of 6OBAC are difficult to distinguish, if simply observed at the microscope. There are only slight changes in the texture from a crystal texture to the other. The image processing, previously developed for smooth transitions, proves effective to determine a precise value of the transition temperatures. The same is valid for the study of the nematic phase and for evaluating the texture transition.

In spite of the absence of a smectic phase, the 6OBAC exhibits two sub-phases with different textures, as it happens in other materials of the homologue series which possess the smectic phase. This is the first material that exhibits a texture transition in a nematic phase, originated directly from a crystal phase. Moreover, 6OBAC has a behaviour of polarisation colours, near the clearing point, very similar to that of hexylcyclohexane carboxylic C6 acid. This suggests that 6OBAC has oligomers too near the clearing point.

**Appendix**

Images are detected by a CCD camera and then stored and elaborated on a workstation. We consider grey-level images $g(x,y)$ with pixels $l_x \times l_y$. The mean intensity $M_o$ and the $k-$rank statistical moments:

$$M_o = \frac{1}{l_x l_y} \sum_{x=0}^{l_x} \sum_{y=0}^{l_y} g(x,y) \quad (1)$$

$$M_k = \frac{1}{l_x l_y} \sum_{x=0}^{l_x} \sum_{y=0}^{l_y} [g(x,y) - M_o]^k \quad (2)$$

of the $(x,y)$ rectangular image frame are evaluated. The second order statistical moment is the variance of the image. Starting from an arbitrary point, in a chosen direction, for instance the $x-$direction, we calculate the mean value of pixel grey tones over a distance $l_0$:

$$M_o^x(x,y) = \frac{1}{l_o} \sum_{\mathbf{x}=0}^{l_o} g(x+\mathbf{x}, y) \quad (3)$$

Distance $l_o = l_o^x(x,y)$ is function of the starting point $P(x,y)$ and defined as the distance where the value of directional moment $M_o^x(x,y)$ reaches, within a threshold level $t$, the value of image moment $M_o$: function $l_o^x(x,y)$ then represents the behaviour of local coherence lengths relevant to the mean intensity. We choose to define these functions as coherence lengths according to their actual use in physics. For instance, coherence length in optics is the propagation distance from a coherent source to a point where an electromagnetic wave maintains a specified degree of coherence. In the liquid crystal physics, coherence length measures the distance where order parameter maintains itself almost constant. Averaging on the whole image frame,

$$L_o^x = \frac{1}{l_x l_y} \sum_{x=0}^{l_x} \sum_{y=0}^{l_y} l_o^x(x,y) \quad (4)$$

the mean coherence length $L_o^x$ is obtained. The same procedure is applied in $y$-direction for coherence $L_o^y$. Coherence lengths in all the directions of the image plane (not only $x-$ and $y-$directions), provide a polar diagram as shown in Fig.3-4. From the polar diagram, texture dimensions and texture anisotropy receive quantitative evaluations.

Let us note that the coherence length evaluation is not depending on the value of mean intensity $M_o$ and on image contrast. Threshold $t$ is usually chosen as the ratio $s/M_o$ where $s$ is the standard deviation of the image grey tones, obtained by the second order moment.

TABLE I

| Phase transition | Temperature (°C) |
| --- | --- |
| Cr-I – Cr-II | 49 |
| Cr-II – Cr-III | 68 |
| Cr-III – Cr-IV | 93 |
| Cr-IV – N1 | 106 |
| N1 – N2 | 129 |
| N2 – Isotropic phase | 147 |

**Table I: Transition temperatures in 6OBAC. In agreement with Ref. [16]**

**FIGURE CAPTIONS.**

Fig.1 Closed and open dimers with hydrogen bonds.

Fig. 2 The four crystalline modifications of 6OBAC, detected with optical microscopy. Images are recorded at 45°C (I), 60°C (II), 75°C (III) and 100°C (IV). The size of the image is 1 mm.

Fig. 3 Transition from the nematic sub-phase N1 to N2 in 6OBAC.

Fig.4 Behaviour of mean intensity M and of variance V obtained from the image maps as a function of temperature. The mean intensity is not able to show the transitions, but the variance curve has discontinuities. Curve D represents the difference between image frames in the scanning sequence and is very sensitive to texture transitions.

Fig. 5 Polar diagrams showing shape and size of the domains, defined as in Ref.12, for image frames of the four crystalline modifications, shown in Fig.2.

Fig. 6 Mosaic texture with striped domains of the Cr-IV in 6OBAC.

Fig.7 8OBAC has crystal-crystal modification. On the left, the crystal at low temperature and on the right, the crystal preceding the smectic phase. Note the stripes in the texture as those seen in 6OBAC.

Fig.8 Transition form Cr-IV to nematic: some domain boundaries remain in the nematic phase. The polar diagrams confirm the memory of the domains in the nematic phase. The units on the axes represent number of pixels.

Fig.9 A model to explain the origin of different textures in the nematic phase. From a smectic C structure with schlieren defects, the material passes in the nematic phase N1. In some points of the walls, smectic planes remain anchored and there cybotactic clusters maintain the local smectic order. When the temperature further increases (sub-phase N2), the smectic order in cybotactic cluster is suppressed.

Fig.10 In the sub-phase N2, 15 degrees before the clearing point, the domain colour passes from green to red and then to green again, a behaviour previously observed in C6. On the right side the maps of the histograms for the three colours tones is proposed.

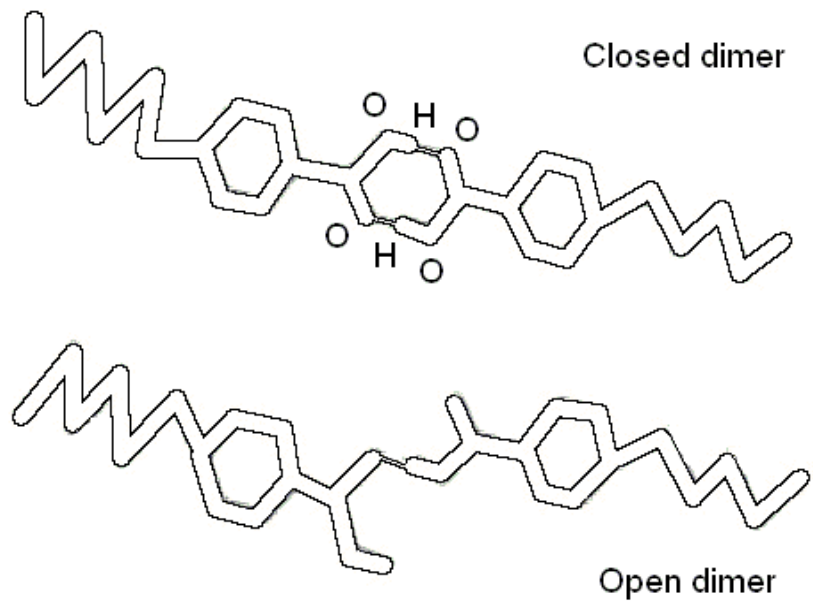

**FIGURE 1**

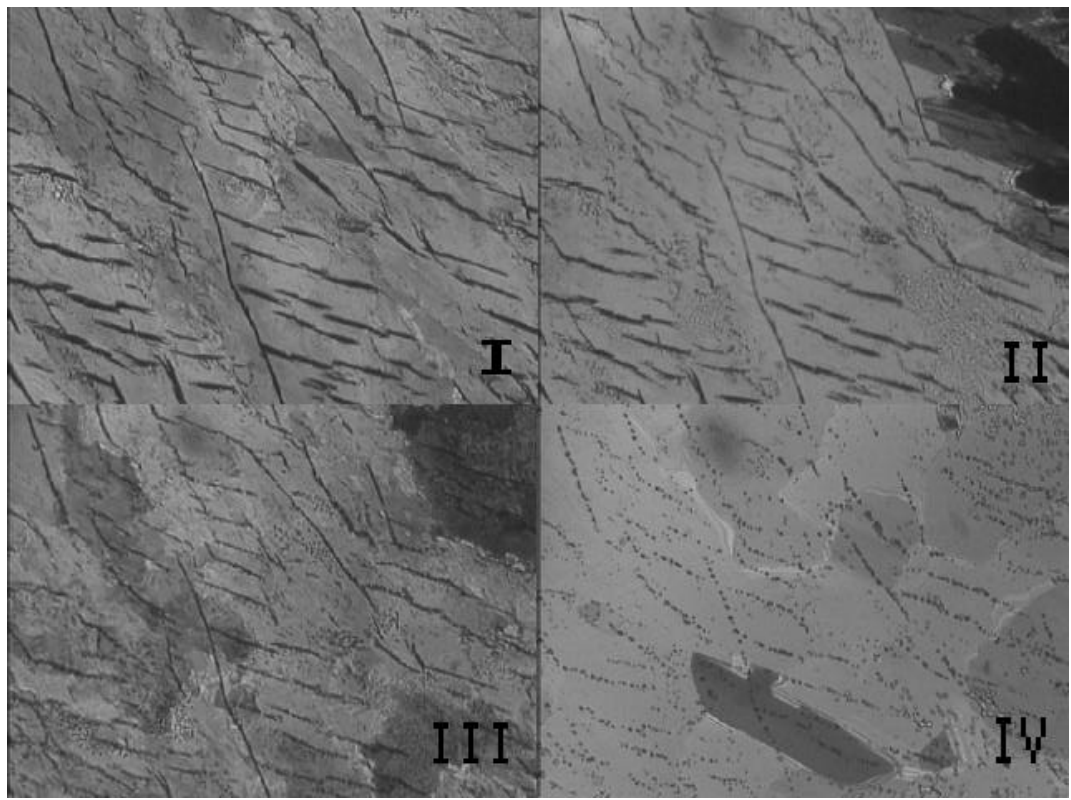

**FIGURE 2**

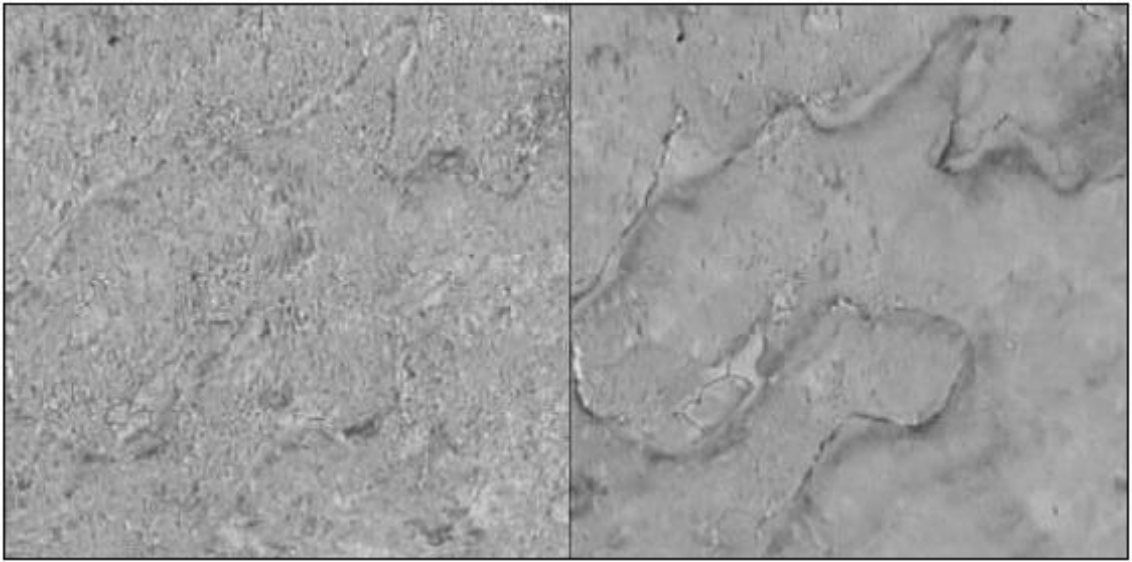

**FIGURE 3**

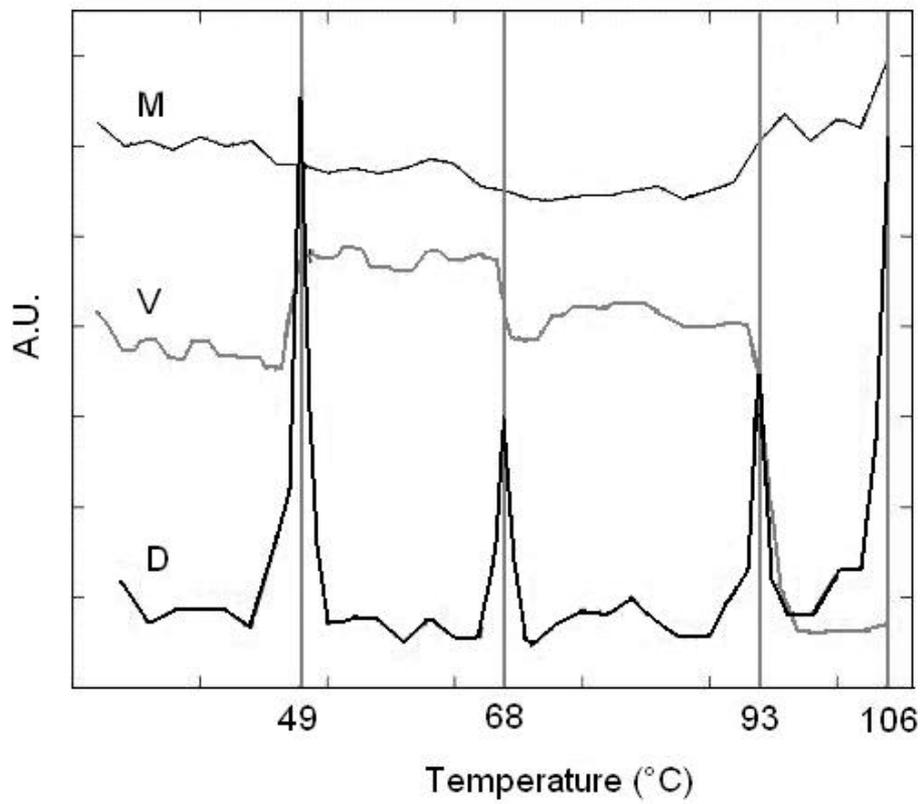

**FIGURE 4**

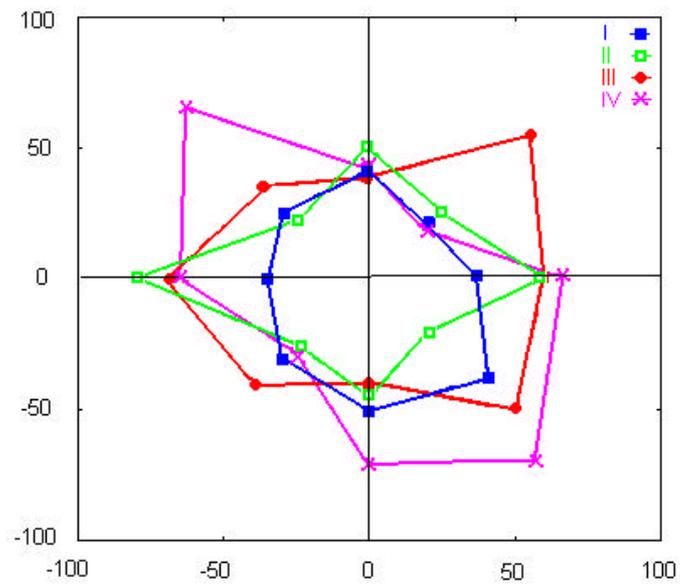

**FIGURE 5**

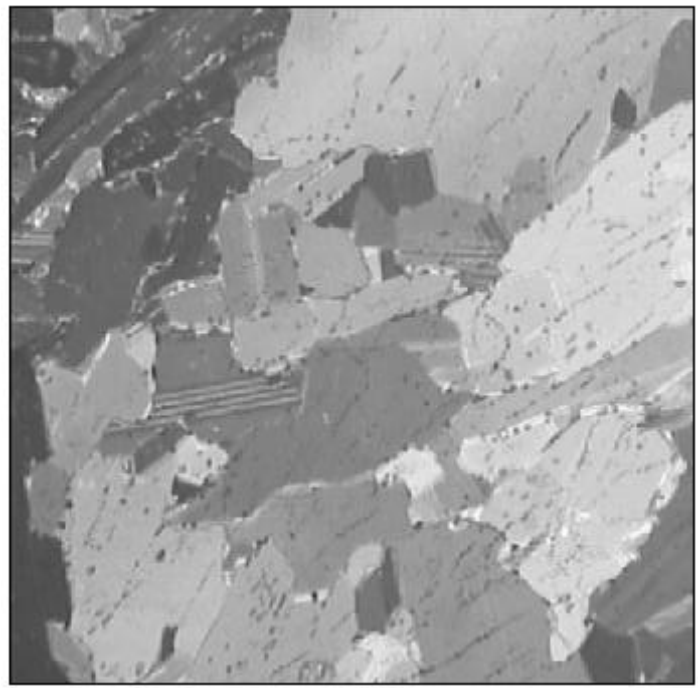

**FIGURE 6**

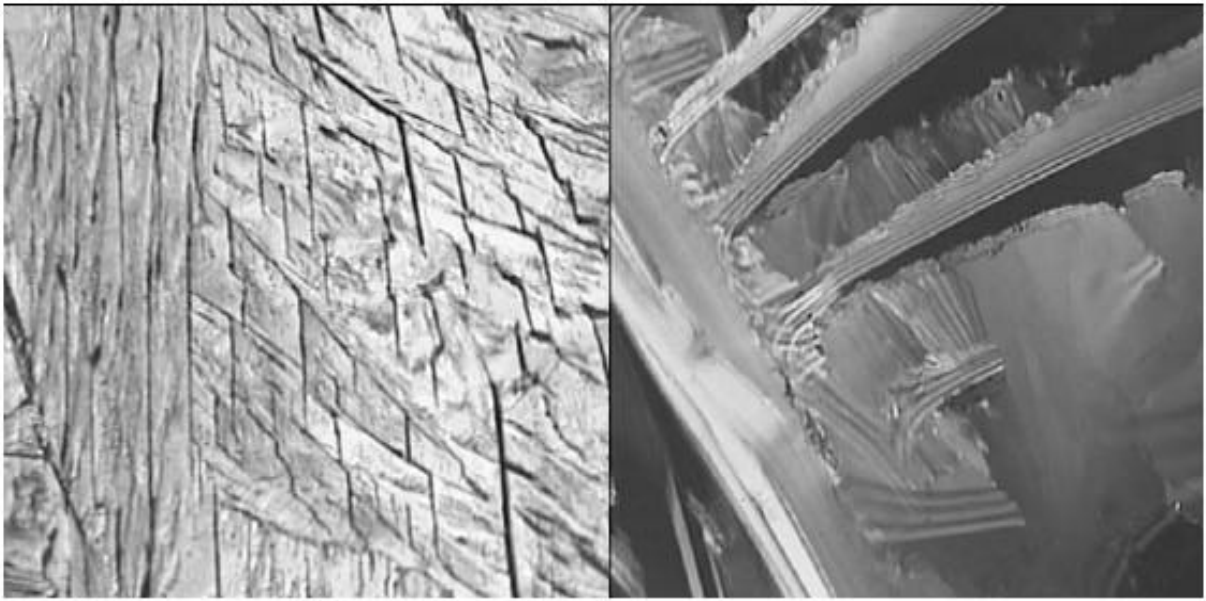

**FIGURE 7**

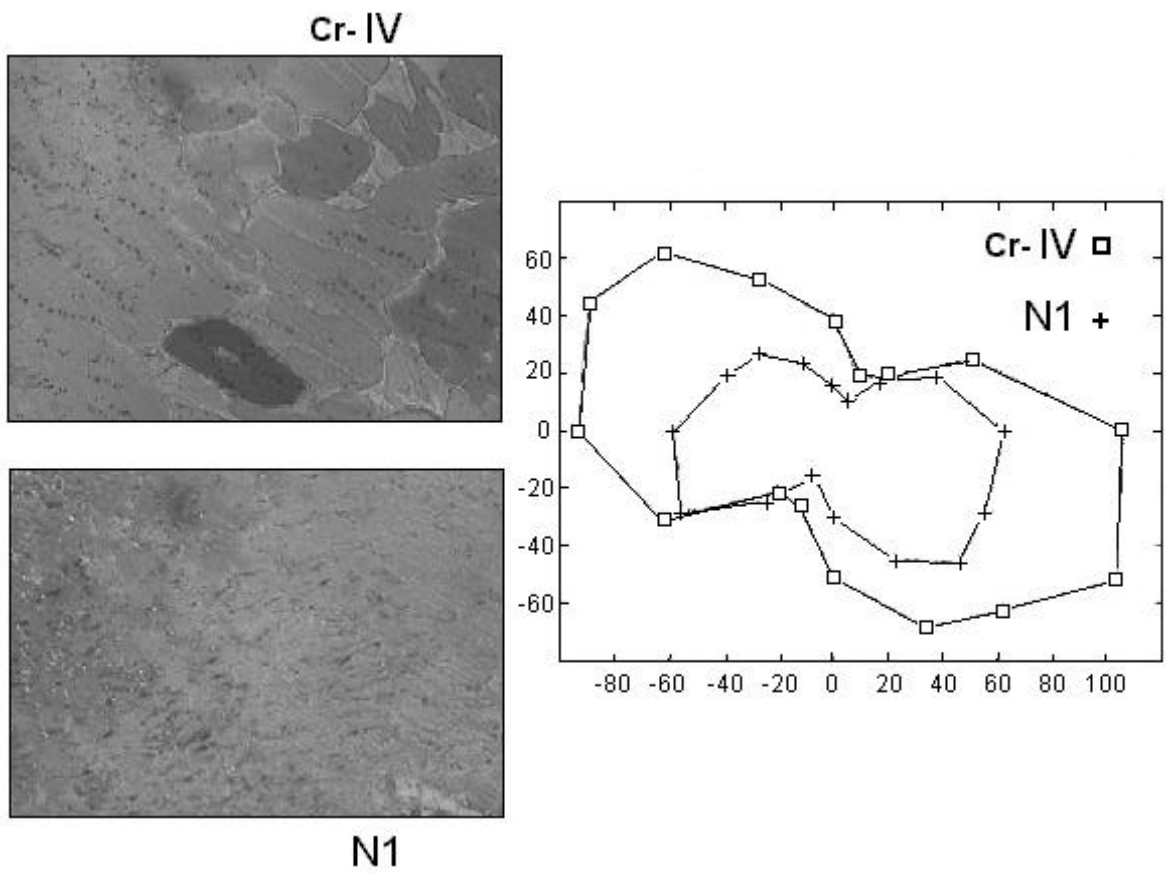

**FIGURE 8**

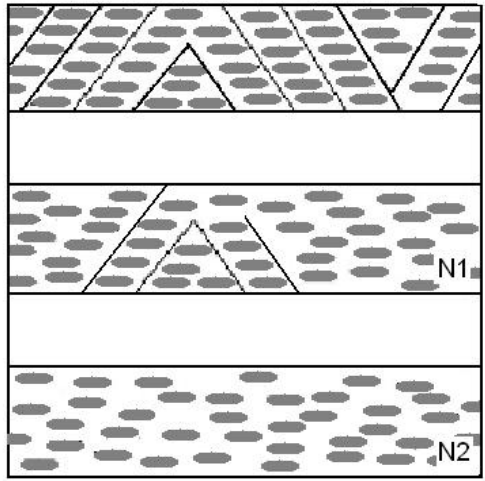

**FIGURE 9**

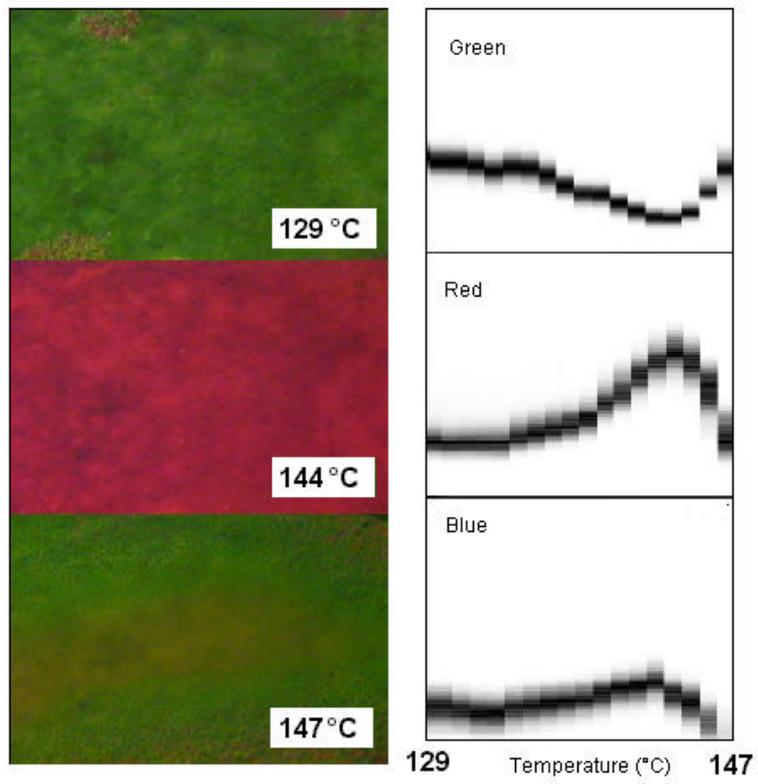

**FIGURE 10**